\documentclass[12pt]{article}
\usepackage[dvips]{graphicx}
\usepackage{epsfig}
\usepackage{textcomp}
\usepackage{amsmath}
\usepackage{amssymb}
\usepackage{bm}
\def\slashed#1{\kern+0.1em /\kern-0.65em #1}

\newcounter{affno}
\renewcommand\theaffno{\alph{affno}}
\newcommand{\aff}[1]{\refstepcounter{affno}\label{#1} \textsuperscript{\theaffno}}
\newcommand{\aref}[1]{\textsuperscript{\ref{#1}}}
\newcommand{\arefs}[2]{\textsuperscript{\ref{#1},\ref{#2}}} 

\begin{document}

\title{Discovering new physics in rare kaon decays}

\date{March 21, 2022}

\author{Thomas Blum\aref{UConn},
Peter Boyle\arefs{BNL}{UoE},
Mattia Bruno\arefs{INFN}{Milan}, 
Norman Christ\aref{CU}, \\
Felix Erben\aref{UoE},
Xu Feng\aref{BU},
Vera Guelpers\aref{UoE},
Ryan Hill\aref{UoE}, \\
Raoul Hodgson\aref{UoE},
Danel Hoying\aref{BNL}, 
Taku Izubuchi\arefs{RBRC}{BNL}, 
Yong-Chull Jang\aref{CU}, \\ 
Luchang Jin\aref{UConn}, 
Chulwoo Jung\aref{BNL},
Joe Karpie\aref{CU}, 
Christopher Kelly\aref{BNL}, \\ 
Christoph Lehner\aref{UR},  
Antonin Portelli\aref{UoE}, 
Christopher Sachrajda\aref{Soton}, 
Amarjit Soni\aref{BU}, \\
Masaaki Tomii\aref{UConn}, 
Bigeng Wang\arefs{CU}{KU} and Tianle Wang\arefs{CU}{BNL} \\
(RBC and UKQCD collaborations)} 

\maketitle

\begin{abstract}
The decays and mixing of $K$ mesons are remarkably sensitive to the weak interactions of quarks and leptons at high energies.  They provide important tests of the standard model at both first and second order in the Fermi constant $G_F$ and offer a window into possible new phenomena at energies as high as 1,000 TeV.  These possibilities become even more compelling as the growing capabilities of lattice QCD make high-precision standard model predictions possible.  Here we discuss and attempt to forecast some of these capabilities.
\end{abstract}

The increasing precision and expanding reach of lattice QCD calculations create new opportunities to search for  phenomena that lie outside of the standard model.  With more accurate standard model predictions, previous experimental results acquire greater sensitivity to beyond-the-standard-model phenomena at higher energies and with weaker couplings.   These new theoretical capabilities may also motivate future experiments if the results from these experiments can be predicted with greater accuracy from the standard model.  In this white paper we focus on low-energy phenomena involving kaon mixing and decay including both first- and second-order weak processes.  This is the region in which current lattice calculations are most accurate with well-understood systematic errors and where the barriers to obtaining predictions with substantially smaller errors appear surmountable.   In the sections below we describe the current state-of-the-art for the lattice QCD predictions of a number of such processes and discuss the opportunities for significant improvement over the next decade.

\section{Direct $CP$ violation and $\varepsilon'$}
\label{sec:epsilon-prime}

The standard model predicts $CP$ violation in $K_L\to\pi\pi$ decay, both through explicit decay matrix elements, $\langle\pi\pi|H_W|K_L\rangle$, with complex $CP$-violating phases (direct $CP$ violation described by the parameter $\varepsilon'$) and though $CP$-violating mixing between the $CP$ eigenstates $(K^0 \pm \bar K^0)/\sqrt{2}$ (indirect $CP$ violation described by the parameter $\varepsilon$).  The Kobayashi-Maskawa theory of $CP$ violation in the standard model predicts a very small value for $\varepsilon'$ because of the requirement that all three families of quarks must contribute to the amplitude.  The parameter $\varepsilon'$ was determined in challenging measurements by the NA28 and KTeV experiments and is usually presented as the ratio Re$(\varepsilon'/\varepsilon) = 16.6(2.3)\times 10^{-4}$~\cite{ParticleDataGroup:2018ovx}.  Since the indirect $CP$ violation parameter $\varepsilon = 2.228(11) \times 10^{-3}$ is itself very small, these direct effects are on the order of one part per million, implying that comparison of experiment with the standard model predictions offers a highly sensitive opportunity to discover the presence of new sources of $CP$ violation.

Calculations of $\varepsilon'$ in the standard model from first principles have become available only recently~\cite{RBC:2015gro, Abbott:2020hxn} with a result Re$(\varepsilon'/\varepsilon) = 21.7(2.6)(6.2)(5.0)\times 10^{-4}$ that is consistent with experiment.  Here the left-most errors are statistical and systematic while the third error estimates the omitted effects of electromagnetism and the difference between the up and down quark masses.  This lattice QCD result is the fruit of a more than 25-year effort that depends on a number of important advances in method, for example a lattice fermion formulation (DWF) with tightly controlled violations of chiral symmetry~\cite{Kaplan:1992bt, Shamir:1993zy},  the ability to treat two-pion states in finite volume~\cite{Lellouch:2000pv} and a method (RI/SMOM) to accurately renormalize composite lattice operators with continuum conventions~\cite{Martinelli:1994ty, Sturm:2009kb}.  

The increasingly accurate calculation of $\varepsilon'$ remains one of the most important goals of the RBC and UKQCD collaborations.  We are now carrying out calculations following two complementary computational strategies.   The first is a continuation of our previous work in which G-parity boundary conditions were imposed on the quark and gauge fields.  This results in pions which obey anti-periodic boundary conditions so that the first finite-volume state with energy above the vacuum is a two-pion state with non-zero pion relative momentum whose energy can be tuned to equal the kaon mass giving on-shell decay matrix elements~\cite{Lellouch:2000pv}.  Collecting large statistics and using a bootstrap method to determine the appropriate $\chi^2$ distribution for correlated fits~\cite{Kelly:2019wfj} allows multi-parameter fits which incorporate nearby $\pi\pi$ excited states and give meaningful $p$-values for the resulting fits.

The second approach being developed uses standard periodic boundary conditions.  With periodic boundary conditions $\pi\pi$ states with near-zero relative momenta are allowed and the energy-conserving decay matrix element must then involve an excited, higher-energy finite-volume $\pi\pi$ state which the Generalized Eigenvalue Problem (GEVP) method~\cite{Luscher:1990ck, Blossier:2009kd} is used to identify~\cite{Tomii:2022vxb}.  This quite different method should provide important verification of the earlier G-parity results and remove the burden of creating special G-parity ensembles of gauge configurations.   A further advantage is that the use of periodic boundary conditions will allow the introduction of electromagnetism which is obstructed by the charge non-conserving character of G-parity boundary conditions which transform an up quark into an anti-down quark as it passes through the boundary.

We will briefly discuss the major sources of error in the most recent lattice result for $\varepsilon'$~\cite{Abbott:2020hxn} and the prospects for their reduction.  We will provide necessarily uncertain estimates of the computer resources needed for such future calculations based on the approximately 500 M Cori core hours or 1.5 Exaflops-hours ($3600\times10^{18}$ floating-point operations) required for the results presented in Refs.~\cite{Abbott:2020hxn, RBC:2021acc}.

\begin{enumerate}

\item \underline{Statistics.}  The calculation of the $I=0$ component of $K\to\pi\pi$ decay is made difficult by the vacuum quantum numbers of the $I=0$ state and the resulting disconnected diagrams.  The 10\% statistical error for the result in Ref.~\cite{Abbott:2020hxn} was achieved with 741 Monte Carlo samples which is relatively large for a lattice QCD calculation.  While demanding of computer resources, increasing statistics poses no fundamental barrier and this error can in principle be reduced to whatever level is needed.

\item \underline{Finite lattice spacing.}  The current calculation of $\varepsilon'$ is performed using a single lattice spacing and the resulting errors estimated from the calculation of the $I=2$ decay amplitude $A_2$ whose lattice QCD calculation~\cite{Blum:2015ywa} includes a continuum extrapolation.  The RBC and UKQCD collaborations have begun a long-term project to extend the current calculation with G-parity boundary condition to two smaller lattice spacings with first calculations underway on the Perlmutter machine at NERSC.   These calculations have the prospect of removing this source of systematic error at the cost of an increased statistical error from the necessary $a\to0$ extrapolation.  This initial study of finite lattice spacing errors plans to compare the original $32^3\times64$ lattice volume with inverse lattice spacing $1/a=1.38$~GeV with two finer ensembles: ($40^3\times 96$, $1/a=1.7$~GeV) and ($48^3\times 96$, $1/a = 2.1$~GeV) at computational costs of 1.7 and 2.1 Exaflops-hours respectively.

It should be noted that the continuum extrapolation carried out in Ref~\cite{Blum:2015ywa} involved a 10\% change in value for Im($A_2$) so that performing such an extrapolation for $\varepsilon'$ is important.  Achieving percent-level errors will require an extrapolation involving more than two values for the lattice spacing and a comparison with descriptions of the lattice spacing dependence which include more than a single $a^2$ term.  

Such a more accurate $a\to0$ extrapolation should also be possible using G-parity versions of current Iwasaki-action ensembles with but with somewhat smaller values of the lattice spacing: ($48^3\times96$, $1/a=2.1$~GeV), ($64^3\times128$, $1/a=2.8$~GeV) and ($96^3\times192$, $1/a=4.1$~GeV) at a cost of 7, 23 and 120 Exaflops-hours respectively.  Here generation of the ($96^3\times192$, $1/a=4.1$~GeV) ensemble would require techniques not yet proven to reduce critical slowing down.  However, such an extensive continuum-limit study may be more appropriate when other errors have been reduced and should likely include an active charm quark and even smaller values of perhaps four different lattice spacings.
 
\item \underline{Short-distance effects.}  As with all lattice QCD calculations of weak interaction phenomena, the predictions of the standard model are provided by a four-fermion effective weak Hamiltonian which contains Wilson coefficients that are evaluated to first- or second-order in electroweak perturbation theory and to some finite order in QCD perturbation theory~\cite{Buchalla:1995vs}.  We should distinguish two important possibilities.  First, the effective weak Hamiltonian includes the effects of virtual charm quarks and is to be used in a three-flavor QCD calculation in which only $u$, $d$  and $s$ quarks appear: the charm quark has been ``integrated out''.  This is the case for the $\varepsilon'$ calculation presented in Ref.~\cite{Abbott:2020hxn}.  

Such a three-flavor effective theory contains two systematic errors arising from this treatment of short distance physics.  By treating the charm quark as very heavy, terms of order $(\Lambda_{\mathrm{QCD}}/m_c)^2$ are being ignored.  These errors are often described as resulting from omitted higher-dimension operators which would have Wilson coefficients with an extra factor of $(1/m_c)^2$.  While we might hope that such effects are on the order of a few percent, a four-flavor calculation is needed if these errors are to be known and eliminated.

The second source of systematic error arises because finite-order QCD perturbation theory is used to calculate the Wilson coefficients.  These essential perturbative calculations relate the four-fermion effective weak Hamiltonian to the actual standard model prediction obtained from the exchange of $Z$'s and $W$'s.  For the three-flavor case, the perturbative calculation of the Wilson coefficient must involve energy scales at or below the charm quark mass, a relatively low energy region in which the convergence of QCD perturbation theory may be uncertain.  The resulting uncertainty in the Wilson coefficients is an important source of error in the $\varepsilon'$ result in Ref.~\cite{Abbott:2020hxn}. 

This QCD perturbation theory error in a three-flavor calculation might be removed by introducing non-perturbative methods to relate the three-flavor theory with the more accurate theory with four quark flavors.   Rather than performing the entire $K\to\pi\pi$ calculation with four flavors, the Wilson coefficients alone might be determined non-perturbatively by working in a four-flavor theory but in a smaller volume without physical pions to match non-perturbative Green's functions involving three- and four-flavor operators with external states that do not involve charm quarks.  Developing a method to do this is an active research project~\cite{Tomii:2020smd} and promises to eliminate almost all short-distance error from a three-flavor calculation of $\varepsilon'$ except that arising from higher dimensional operators.   

With this approach both the three-flavor calculation and four-flavor calculation discussed below would rely on the accurate perturbative calculation of the four-flavor Wilson coefficients.  These errors can be reduced by extending the usual non-perturbative RI/SMOM to a higher energy scale, increasing the lowest energy at which QCD perturbation theory is used and by working to higher order in QCD perturbation theory.  The Wilson coefficients used in the calculation of $\varepsilon'$ in Ref.~\cite{Abbott:2020hxn} relied on one-loop QCD perturbation theory~\cite{Buchalla:1995vs}.   A very important advance in this topic~\cite{Aebischer:2019mtr} will be a two-loop calculation~\cite{Cerda-Sevilla:2018hjk} that may be nearing completion.  This combination of working at a higher scale and including two-loop renormalization results promises to reduce the errors in the Wilson coefficients to the percent level.  Even greater accuracy could be achieved if needed.

A more accurate treatment would simply include the charm quark in the lattice calculation and use a four-flavor effective Hamiltonian which includes pairs of operators creating and destroying the charm quark.  Of course, the resulting four-flavor lattice calculation would be much more computationally costly than the corresponding three-flavor version.  In addition to the problems from disconnected graphs and identifying the energy-conserving $I=0$ $\pi\pi$ final state, one must accurately treat a virtual charm quark loop.  The charm quark mass introduces a new, higher energy scale which  requires working at substantially smaller lattice spacing while maintaining the same size physical volume as was needed for the three-flavor calculations.  In fact, in such a four-flavor calculation the virtual charm quark's momentum is on the order of the charm mass, increasing the requirement of small lattice spacing beyond that needed for the more common treatment of valence charm quarks in $D$ or $D^*$ mesons where the charm quark momentum is on the order of $\Lambda_{\mathrm{QCD}}$, considerably smaller than the charm quark mass and allowing a variety of heavy-quark effective theories to be used.   

A realistic four-flavor calculation of $\varepsilon'$ has not yet been attempted and is likely five or more years in the future.  However, when adequate resources become available this will be an important improvement to make and does not pose any fundamental difficulties.   Such a four-flavor calculation should be able to reduce the errors from the truncation of QCD perturbation theory and the omission of higher dimension operators (which would now arise from the bottom quark) to the percent level.  Based on our present experience including a charm valence quark in the calculation of $\Delta M_K$ described below, such a four-flavor calculation would require significantly smaller lattice spacings than presently available.   Four-flavor calculations of $\varepsilon'$ on three ensembles  ($64^3\times256$, $1/a=2.76$~GeV), ($96^3\times384$, $1/a=4.14$~GeV) and ($128^3\times512$, $1/a=5.51$~GeV) at a cost of 50, 240 and 1500 Exaflops-hours respectively.   (The time extent of each lattice volume has been doubled to allow for the use of open boundary conditions~\cite{Luscher:2011kk}, a step that may be needed to control finite-volume errors associated with frozen topology.)

\item \underline{Isospin-violating corrections.}  For most quantities a careful treatment of isospin violating effects becomes important only when one wishes to reduce systematic errors well below the level of 1\%.  However, as is well known, the facts that indirect $CP$ violation in $K_L\to\pi\pi$ decay requires non-zero values for both the $I=0$ and $I=2$ amplitudes $A_0$ and $A_2$ and that $A_2$ is roughly twenty times smaller than $A_0$ implies that isospin breaking corrections to $A_0$ will produce corrections to $A_2$ which are effectively twenty times larger than they would typically be.  

A thorough study using the large $N_c$ approximation and chiral perturbation theory~\cite{Cirigliano:2019cpi} found these isospin breaking contributions to $\varepsilon'$ to be 25\% with a nearly 100\% uncertainty.  Making a first-principles calculation of this correction which verifies and refines the result of Ref.~\cite{Cirigliano:2019cpi} is now a high priority goal of lattice QCD.  Such a calculation is complicated both by the three weak and electromagnetic vertices that appear in the decay amplitude to be computed and by the long-range character of the electromagnetic interaction.

Current efforts introduce Coulomb gauge to separate the calculation into two independent components.  The first is the correction resulting from the instantaneous Coulomb interaction between the quark charges.  This source of isospin breaking appears amenable to the methods of lattice QCD allowing a treatment of finite volume effects with errors which decrease exponentially as the linear extent of the lattice increases~\cite{Christ:2021guf}.  An exploratory calculation is underway using this method to study the electromagnetic corrections to $\pi^+\pi^+$ scattering.  With the introduction of isospin breaking effects, the $I=0$ and $I=2$ channels mix.  Fortunately it appears~\cite{Christ:2017pze} that the resulting two-channel problem can be treated using lattice QCD in finite volume by existing techniques~\cite{He:2005ey, Hansen:2012tf} combined with exploiting perturbation theory in $\alpha_{\mathrm{EM}}$.  (The effects of the isospin breaking mass difference between the up and down quarks can be included in such a calculation without added difficulty.)

The second component of such a lattice QCD calculation of the isospin breaking corrections to $\varepsilon'$ is less well understood.  Here the effects of the transverse photons present in Coulomb gauge must be computed.  While the effects of soft radiation can be accurately treated classically, more energetic virtual photons which are sensitive to hadronic structure require lattice QCD and new methods may need to be developed to deal with the effect of the three-body $K\to\pi\pi\gamma$ decay.

\item \underline{Finite-volume errors}  In most lattice QCD calculations finite-volume errors can be estimated by simply comparing results from two or more physical volumes.  However, the energy-conserving condition on the finite-volume $\pi\pi$ energy, $E_{\pi\pi}=M_K$ in the calculation of $\varepsilon'$ makes a change of physical volume more difficult.  The finite-volume errors can be divided into two classes.  The first are errors arising from approximations made when the Lellouch-L\"uscher finite-volume treatment of this decay is implemented~\cite{Lellouch:2000pv}.  These are corrections which fall as inverse powers of the physical volume and can likely be controlled at the percent-level by careful analytic study of the terms in the Lellouch-L\"uscher treatment which have been neglected.  This was done, for example, in Section VII.E. {\it Higher partial wave correction} of Ref.~\cite{RBC:2021acc}.   

The second class of finite-volume error arises from the bound states and the decay process being ``compressed'' into a finite volume and decreases exponentially as the linear size of the system increases.  These errors can be estimated using chiral perturbation theory.  In our recent calculation~\cite{Abbott:2020hxn} this source of finite-volume error is estimated at 7\% from an inflated value of the finite-volume error suggested by chiral perturbation theory applied to the amplitude $A_2$.  To be confident of such error estimates at the 1\% level, it will likely be necessary to perform calculations on two or more volumes, working with what will be multiple excited finite-volume states with $E_{\pi\pi}=M_K$.  Here the $K\to\pi\pi$ project with periodic boundary conditions is already developing the techniques that will be needed.

\end{enumerate}

As discussed above there are substantial theoretical and computational obstacles to a lattice QCD calculation of the standard model prediction for $\varepsilon'$ with reduced errors.  However, it may not be unreasonable to expect that with continued effort a reduction in errors below the 30\% level in five years and below 10\% in ten years may be achieved.

\section{The $K_L-K_S$ mass difference}
\label{sec:DeltaMK}

With a measured value of $3.482(6)\times10^{-12}$ MeV this small mass difference ($\Delta M_K$) has sensitivity to physics at the 1,000 TeV scale and yet has been predicted by the standard model with only 36\% accuracy~\cite{Brod:2011ty}.  The charm quark and the GIM mechanism play a critical role in determining $\Delta M_K$ in the standard model.  In fact the small size of $\Delta M_K$ may have been the first evidence for the existence of the charm mass scale~\cite{Mohapatra:1968zz}.  Using a three-flavor weak Hamiltonian in second-order electro-weak perturbation theory results in a quadratically divergent expression for $\Delta M_K$.  Moving to the four-flavor theory this quadratic divergence is removed and the resulting expression is finite, provided the small effects of the top quark are neglected and the high-momentum GIM cancellation between the charm and up quarks is treated as exact.  

Thus, to few-percent accuracy, all contributions to $\Delta M_K$ come from energy scales between $\Lambda_\mathrm{QCD}$ and $m_c$.  This suggests that the methods of lattice QCD may be sufficient for a complete calculation of $\Delta M_K$, provided a sufficiently small lattice spacing is used to accurately treat the virtual charm quark loops.  This is in fact the case -- making the accurate calculation of $\Delta M_K$ important opportunity for lattice QCD~\cite{Christ:2010gi, Christ:2012se, Bai:2014cva, Christ:2015pwa, Wang:2021twm}.  

Here we discuss the present state of our on-going calculation of $\Delta M_K$ using physical quark masses on a $64^3\times128$ lattice volume with inverse lattice spacing of $1/a=2.38$~GeV~\cite{Wang:2021twm} and the longer-term prospects for future calculations at smaller lattice spacing.  The preliminary result from this calculation is $\Delta M_K = 5.8(0.6)_\mathrm{stat}(2.3)_\mathrm{sys}\times10^{-12}$~MeV.  This calculation was performed using 152 gauge configurations and required approximately 200 M core hours on the Mira machine at the Argonne National Laboratory.  This corresponds to a total number of floating point operations of 0.5 Exaflops-hours.  We now discuss the various errors present in the current calculation and the prospects for reducing them.

\begin{enumerate}
\item \underline{Statistics}  The calculation of $\Delta M_K$ is carried out by studying the Euclidean-space Green's function 
\begin{equation}
\int_{t_i+T}^{t_f-T}dt_2\int_{t_i+T}^{t_f-T}dt_1\left\langle \left(K^0(t_f)\right)^\dagger H_W(t_2) H_W(t_1) K^0(t_i).\right\rangle
\label{eq:DeltaMK}
\end{equation}
Here $H_W$ is the $\Delta S=-1$ effective weak Hamiltonian, $K^0$ an interpolating operator which creates a kaon and $T$ is chosen sufficiently large that only the matrix element with an initial $K^0$ and final $\overline{K}^0$ state appears.  While this treatment of the initial and final kaon states poses no special difficulties, the possible intermediate states which can occur between the two weak operators present serious challenges.  A Euclidean Green's function such as that in Eq.~\eqref{eq:DeltaMK} contains a number of distinct transition amplitudes which must be identified and separated.  Intermediate vacuum, single-pion, two-pion and $\eta$ states contribute to both $\Delta M_K$ as well as to terms growing exponentially in the separation $|t_2-t_1|$ relative to the contribution which gives $\Delta M_K$ (or falling slowly with increasing $|t_2-t_1|$ as is the case for the $\eta$).  These latter contributions must all be removed and for states much lighter than $M_K$ large statistical noise remains after the unwanted contributions of these states have been subtracted.  

Dealing with the contribution of these unwanted transition amplitudes is especially challenging for the $\eta$ in which statistically noisy disconnected diagrams play a large role and where the near degeneracy of the $\eta$ and the kaon results in a near-vanishing energy denominator.  While still a significant source of statistical error, the physical and unphysical contributions of the $\eta$ are best simply eliminated by introducing a physically-irrelevant $\bar s d$ term into $H_W$ with its coefficient adjusted to achieve $\langle \eta|H_W|K^0\rangle=0$.

As suggested by the 10\% statistical error given above, these statistical problems have been largely solved.  The analysis of an increased number of Monte Carlo samples, a careful treatment of the integral over the relative time separation $t_2-t_1$ in Eq.~\eqref{eq:DeltaMK} known as the ``single-integration method'' to avoid integration regions that contribute only noise and a shift in computational strategy to devote the largest number of samples to the computation of only those diagrams with the largest statistical noise has reduced the original statistical errors~\cite{Bai:2018mdv} three fold.  It is reasonable to expect a further $3\times$ to $10\times$ reduction in statistical errors over the next decade driven by the increase in computer capability and continued algorithmic ingenuity. 																															\item \underline{Finite lattice spacing.}  At present the problem of non-zero lattice spacing is the largest difficulty faced by the lattice calculation of $\Delta M_K$ and is reflected in the large systematic error in our current preliminary result for $\Delta M_K$~\cite{Wang:2021twm} given above.  In fact, the completion of further studies of this systematic error is the final step in our physical-mass calculation of $\Delta M_K$~\cite{Wang:2021twm}.  Our original expectation was that the large charm quark mass would be the dominant source of finite lattice spacing errors in this calculation where the product of bare charm quark mass and lattice spacing $m_c^{(\mathrm{bare})} a \approx 0.3$ is close to the largest value at which the domain wall fermion (DWF) formulation should be expected to be reliable~\cite{Boyle:2017jwu, Tomii:2017lyo}.  To address this issue, a series of five values of $m_c^{(\mathrm{bare})}a$ were studied and no anomalous behavior seen.  

In addition to this exploration of the charm quark mass dependence, we also performed a companion calculation studying the lattice-spacing dependence of some of the amplitudes which contribute to $\Delta M_K$ using two larger lattice spacings, correspondingly smaller charm quark masses and heavier-than-physical light quark masses.  These calculations showed an unexpectedly large, approximate 40\% error coming perhaps equally from the large charm quark mass and $\approx 20$\% scaling violations in the matrix elements of the unfamiliar four-quark operator belonging to the $(20,1)$ representation of $SU_L(4)\times SU_R(4)$ which contributes to $\Delta M_K$.  We note that similar-size scaling violations have been seen in the past for matrix elements of $(8,8)$ operators in studies of the $\Delta I=3/2$ contribution to $K\to\pi\pi$ decay.  However, for that case cancellations reduced the $O(a^2)$ errors in the final physical amplitude~\cite{Blum:2015ywa}.

This experience suggests that a calculation of $\Delta M_K$ with controlled finite lattice spacing errors is best carried out on the soon-to-be-available exascale resources using the three lattice spacings and corresponding lattice volumes also needed for the four-flavor calculation of $\varepsilon'$ discussed above:   ($64^3\times256$, $1/a=2.76$~GeV), ($96^3\times384$, $1/a=4.14$~GeV) and ($128^3\times512$, $1/a=5.51$~GeV) at a cost of 2, 8 and 32 Exaflops-hours respectively.  These estimates include a four times increase in statistics with the goal of a 5\% result for $\Delta M_K$ in 2026.

\item \underline{Finite volume errors.}  Because of the contribution of discrete, finite-volume $\pi\pi$ energy eigenstates, a lattice result for $\Delta M_K$ includes potentially large power-law finite-volume effects.  Fortunately the needed finite-volume corrections can be accurately computed and applied by using a generalization of the familiar L\"uscher treatment of $\pi\pi$ scattering in finite volume~\cite{Christ:2015pwa}.  In our present physical-mass calculation~\cite{Wang:2021twm} these corrections are on the order of a few percent.  However, they may become larger depending on the proximity of a quantized $\pi\pi$ energy and the kaon mass.  Should this correction become larger or a greater accuracy desired, more care will be needed when making this correction.  Such an enhanced calculation of this correction would be a natural adjunct to the $K\to\pi\pi$ calculations required to determine $\varepsilon'$.  In the more distant future when a calculation of $\Delta M_K$ with 0.1\% accuracy is attempted it will become important to correct for the more difficult finite-volume effects arising from three-pion states.  This is at present an unsolved problem but one which may have been brought under control~\cite{Hansen:2021ofl} before this degree of precision is needed.

\item \underline{Omitted standard model contributions}  The four-flavor treatment of $\Delta M_K$ described above assumes the upper left $2\times2$ corner of the CKM matrix is unitary and ignores the weak coupling of the top quark to the down and strange quarks.  This introduces a few-percent error into the calculation of $\Delta M_K$~\cite{Christ:2014qwa} -- an error which could be removed by a calculation of the omitted top-quark contribution in QCD perturbation theory.  While such a perturbation theory calculation would itself contain long-distance errors of the sort described below for similar calculations of $\varepsilon$ and $K^+\to\pi^+\nu\bar\nu$, these few-percent corrections would be unimportant until a lattice calculation of $\Delta M_K$ with an accuracy on the order of the experimental result was being attempted.

\item \underline{Other errors.}  The lattice QCD calculation of $\Delta M_K$ will contain a variety of other errors such as those present in the perturbative calculation of the needed Wilson coefficients and finite-volume errors that are exponentially suppressed in the linear lattice size as the lattice volume is increased.  These involve similar considerations to those outlined above in the discussion of $\varepsilon'$.

\end{enumerate}

We conclude that an {\it ab initio}  lattice QCD calculation of $\Delta M_K$ in the standard model which reaches the experimental accuracy is likely not possible within the coming decade.  However, achieving errors on the five-percent level and perhaps smaller can be achieved with adequate computer resources.

\section{Indirect $CP$ violation and $\varepsilon$}
\label{sec:epsilon-LD}

Important lattice QCD calculations of the kaon bag parameter $B_K$ led to early predictions for the short-distance part of $\varepsilon$ which would now be accurate at the one-percent level, except for the larger uncertainties arising from the current errors for $V_{cb}$ -- errors which should substantially decrease over the next decade.  However, to reach sub-percent accuracy, the few-percent contribution of long-distance physics to $\varepsilon$~\cite{Buras:2010pza} must be computed.  This long-distance contribution is closely related to the calculation of $\Delta M_K$ described above.  It is a contribution to the CP-violating, imaginary part of the same off-diagonal element of the $K^0-\overline{K}^0$ mixing matrix whose real part determines $\Delta M_K$.

In contrast to $\Delta M_K$, when computed using the product of two four-flavor operators as in Eq.~\eqref{eq:DeltaMK} this contribution is logarithmically divergent and is made finite only by the structure of the underlying box diagram whose long-distance behavior is represented by the product of these two four-quark operators.  While more complex than the calculation of $\Delta M_K$, this dependence on the short distance structure of the standard model can be accurately handled in a lattice calculation by imposing an RI/SMOM condition to remove the short distance singularity as the two operators in Eq.~\eqref{eq:DeltaMK} approach each other.  In fact an exploratory calculation of this correction to $\varepsilon$ which deals with this difficulty has been carried out~\cite{Christ:2015phf}.  After this renormalization step, the lattice QCD calculation of the long-distance correction to $\varepsilon$ becomes similar to the calculation of $\Delta M_K$ and the discussion in the previous section of errors and computational goals applies.

There are two added complications that should be recognized.  First, the four-quark operators that appear in the $CP$ violating effective weak interaction are more complex than those that are needed to calculate $\Delta M_K$, requiring that more diagrams be evaluated.  This more than doubles the computational cost of these long-distance corrections to $\varepsilon$.  Second, as discussed above this use of the effective weak $\Delta S=1$ Hamiltonian $H_W$ in second order perturbation theory introduces a new logarithmic singularity which can be properly subtracted and replaced by an additional four-quark operator multiplied by a new low-energy constant which encodes the remaining short-distance physics and cannot be determined using lattice QCD.  By using the RI/SMOM scheme to define the short distance behavior of the $H_W H_W$ product, this low energy constant becomes accessible to QCD perturbation theory as the value of a Green's function with four external quark lines carrying large non-exceptional external momenta.  Thus, in order to obtain a physical result for this long-distance correction to $\varepsilon$ this lattice calculation must be augmented by a perturbative calculation of an infrared-safe low energy constant to two or three loops in QCD perturbation theory.  This new low energy constant represents an additional, critical dependence of this lattice QCD calculation on QCD perturbation theory.

\section{ $K^+\to\pi^+\nu\bar\nu$ decay}  

The rare kaon decays $K\to\pi\nu\bar{\nu}$ have attracted increasing interest during the past few decades.  As flavor changing neutral current processes, these decays are highly suppressed in the standard model and thus provide ideal probes for the observation of new physics effects. The known branching ratio measurement~\cite{Artamonov:2008qb} of $K^+\to\pi^+\nu\bar{\nu}$ is almost twice the standard model prediction~\cite{Buras:2015qea}, but with a 60-70\% uncertainty it is still consistent with the standard model. The current experiment, NA62 at CERN, which aims at an observation of $O(100)$ events and a 10\%-precision measurement of $\mathrm{Br}(K^+\to\pi^+\nu\bar{\nu})$, recently reports an upper limit of $1.78\times10^{-10}$ for the $\mathrm{Br}(K^+\to\pi^+\nu\bar\nu)$ at 90\% CL~\cite{CortinaGil:2020vlo}.

These decays are known to be short-distance dominated and are theoretically clean.  The long-distance contributions are safely neglected in $K_L\to\pi^0\nu\bar{\nu}$ and are expected to be small, perhaps of $O(5$\,-\,$10\%)$, in $K^+\to\pi^+\nu\bar{\nu}$ decays.  Though small, by including the long-distance contribution estimated from chiral perturbation theory, the branching ratio $\mathrm{Br}(K^+\to\pi^+\nu\bar{\nu})$ is enhanced by 6\%~\cite{Isidori:2005xm}, which is comparable to the 8\% total standard model error~\cite{Buras:2015qea}.  Since it will be possible to compare the standard model predictions with the new experimental measurement of $\mathrm{Br}(K^+\to\pi^+\nu\bar{\nu})$ relatively soon, a lattice QCD calculation of the long-distance contributions is important and timely.

In the past years, we have developed the theoretical framework~\cite{Christ:2016lro} and performed exploratory numerical calculations with the pion mass approaching its physical value~\cite{Bai:2017fkh,Bai:2018hqu,Christ:2019dxu}. To reach the full physical kinematics, we also need a fine lattice to avoid lattice artifacts arising from the physical mass of the charm quark. We are currently carrying out a physical-kinematics calculation of the long-distance contribution to $K\to\pi\nu\bar{\nu}$ with a target of 30\% total uncertainties.  

This calculation is similar to the calculation of the long distance corrections to $\varepsilon$ and requires a similar RI/SMOM subtraction of a logarithmic dependence on the lattice spacing and a companion perturbative QCD calculation of the corresponding low energy constant.  The discussion of resource requirements and future prospects is thus similar to that found in Sections~\ref{sec:DeltaMK} and \ref{sec:epsilon-LD}.

\section{$K\to\pi\ell^+\ell^-$ decays}

These decays are expected to be extensively observed in the next years through the NA62 and LHCb experiments. They are dominated by standard model long-distance effects featuring an electroproduction of the lepton pair through the intermediate process $K\to \pi\gamma^*$. Using Ward-Takahashi identities, the amplitude of this decay can generally be written
\begin{equation}
  \mathcal{A}_{\mu}[K(k)\to \pi(p)\gamma^*(q)]=-i\frac{G_F}{(4\pi)^2}[q^2(k+p)_{\mu}-(M_K^2-M_\pi^2)q_{\mu}]V(z)\,,
\end{equation}
with $z=q^2/M_K^2$.  A reliable standard model prediction of these decays requires a precise description of the $z$-dependence of the form factor $V(z)$.  A parametrisation commonly accepted~\cite{Cirigliano:2011ny} to describe well the experimental data is
\begin{equation}
  V(z)=a_c+b_cz+V_{\pi\pi}(z)\,,
\end{equation}
where $c\in\{+,S\}$ indicates the kaon state, $a_c$ and $b_c$ are two unknown real constants and $V_{\pi\pi}(z)$ describes the additional contributions above the $q^2=4M_{\pi}^2$ threshold. The constants $a_c$ and $b_c$ are currently only known from experimental measurement~\cite{Batley:2011zz} or phenomenological descriptions~\cite{DAmbrosio:2019xph}. These constants can be correlated to other rare decays, for example in the $B$ sector, through lepton flavour universality violating extensions of the standard model~\cite{Crivellin:2016vjc}. 

Our collaboration pioneered the first calculations of this rare kaon decay in an unphysical context~\cite{Christ:2015aha,Christ:2016mmq}, and recently achieved the first calculation of this amplitude with physical quark masses~\cite{Boyle:2022ccj}. Despite a considerable investment of computing time, we did not manage to obtain a statistically significant result for the $a_+$ parameter at the physical point. This is mainly due to the reduced effect of the GIM cancellation when the up and charm quark have their physical values, leading to a large enhancement of the statistical noise in the diagrams featuring up/charm quark loops. Thanks to this work, we now have a clear direction for
numerically improving our calculation and extending this work is very well suited to the next generation of supercomputers. We believe that over the next 5-10 years, lattice QCD will be in a position to produce predictions of
$a_S,a_+,b_S$, and $b_+$ with uncertainties below the $10\%$ level.

Finally, there are growing efforts in the LHCb experiment~\cite{Aaij:2017ddf,AlvesJunior:2018ldo} to also measure the rare hyperon decay $\Sigma^+\to p\ell^+\ell^-$ which can be seen as a baryon version of the rare kaon decay $K^+\to\pi^+\ell^+\ell^-$. Similarly, it is dominated by long-distance effects through the intermediate process $\Sigma^+\to  p\gamma^*$. Where the rare kaon decays $K^+\to\pi^+\gamma^*$ need only one form factor to describe the amplitude, the decay $\Sigma^+\to p\gamma^*$ needs four form factors because of the higher spin of the external states. The extensive work in Ref.~\cite{He:2005yn} uses various phenomenological approaches to constrain these form factors. Additionally there is a renewed phenomenological interest in these decays in Ref.~\cite{Geng:2021fog}, where they are discussed jointly with the rare kaon decays. We are currently preparing a theoretical publication demonstrating how to compute the rare hyperon amplitude from first principles using lattice QCD. The approach is similar to the one used in the kaon case, but the numerical challenge is expected to be much higher due to the known statistical issues in correlation function involving baryon operators. Beyond supporting the experimental efforts, we are confident that addressing these challenges will push the boundaries of our field for the computation of complex baryon decay processes and the understanding of baryon scattering.

\section{$K_L\to\mu^+\mu^-$ decay}  

This accurately-measured strangeness-changing neutral current process does not allow a corresponding high-profile test of the standard model at second-order because of the background resulting from the largely unknown, two-photon exchange $O(\alpha_{EM}^2 G_F)$ $K_L\to\mu^+\mu^-$ decay amplitude.  Substantial progress over the last few years in using lattice QCD to compute the hadronic light-by-light contribution to the anomalous magnetic moment of the muon~\cite{Blum:2016lnc, Chao:2021tvp} raises the possibility that similar methods may be developed to compute this two photon process which also involves a complex product of QED and QCD amplitudes~\cite{Christ:2020bzb}.  While important problems posed by the calculation of this two-photon exchange amplitude remain unsolved, significant progress has been made toward this goal.  

In particular a complete calculation of the simpler but related decay $\pi^0\to e^+ e^-$ has been carried out with physical quark masses, a continuum limit evaluated and an empirically determined size for both the contribution of disconnected diagrams and the effects of finite volume~\cite{Christ:2020dae}.  The combined statistical and systematic errors on the result are on the order of a few percent.  Of greatest interest may be an analytic method of treating the two-photon intermediate state which eliminates the exponentially growing contribution that would arise from the standard Euclidean-space treatment of two-photon intermediate states with energy below $m_\pi$.  Explicit analytic continuation of the energy in the loop integral involving the two intermediate photons and virtual electron allows the lattice calculation of the entire, complex $\pi^0\to e^+ e^-$ amplitude without exponentially growing terms, with a continuum treatment of the perturbative degrees of freedom and with finite-volume errors that decrease exponentially in the size of the volume within which lattice QCD is used.  

This development effort has now been extended further to the combined weak and electromagnetic decay $K_L\to\gamma\gamma$~\cite{Zhao:2021zzz}.  This exploratory calculation has been carried out using a $24^3\times64$ lattice volume and physical quark masses but with a single, small inverse lattice spacing of $1/a = 1$ GeV.   The hadronic matrix element needed for this decay involves a four-point function with two electromagnetic currents, the effective weak operator $H_W$ and a kaon interpolating operator.   Here unphysical exponential growth can occur as the time elapsed between the absorption of the strange quark by $H_W$ and the emission of the first photon.  As in the calculation of $\Delta M_K$ this relative exponential growth arises from states lighter than the kaon, specifically the $\pi^0$.  An additional unphysical contribution of the $\eta$, while falling exponentially in this separation, falls so slowly that its explicit subtraction is needed.  While the explicit removal of the intermediate $\pi^0$ contribution has been done successfully, a treatment of the $\eta$ similar to what was possible in the calculation of $\Delta M_K$ requires substantially more statistics to yield useful results given the critical role played here by disconnected diagrams.   We expect that a successful treatment of the $\eta$ intermediate state and the associated disconnected diagrams will be possible when this calculation is moved to a larger lattice volume with a smaller lattice spacing because of the increased statistics that will result from more extensive volume averages.  However, the target $K_L\to\mu^+\mu^-$ calculation involves computational challenges that have not yet been solved.  Perhaps the most serious is caused by $\pi\pi\gamma$ intermediate states with energy below $M_K$ where enhanced finite-volume methods are needed.

\section{Conclusion}

All of the phenomena considered here offer significant opportunities to discover new physics in the precision comparison of predictions of the standard model with often difficult-to-obtain experimental results.  In each case the continued development of numerical algorithms and more efficient Euclidean-space lattice field theory methods promise searches for new physics with increased precision.  Each is a challenging lattice calculation which requires substantial access to the most powerful HPC resources and the continuing investment in software and algorithm development needed to make effective use of this cutting edge hardware.

In light of these remarkable advances in theoretical technique and the continued increase in experimental capability, such high-precision studies of rare processes may be ripe for new experiments that exploit our increasing understanding and control of QCD.  The strong interactions among the quarks create unique opportunities for precision experiments but often lead to difficulties in making equally precise predictions -- difficulties which in many cases are now being overcome by the methods of lattice QCD.

\subsection*{\textsf{Affiliations}}
{\small \flushleft
\aff{UConn} Physics Department, University of Connecticut, Storrs, CT 06269-3046, USA \\
\aff{BNL} Brookhaven National Laboratory, Upton, NY 11973, USA\\
\aff{UoE} Higgs Centre for Theoretical Physics, The University of Edinburgh, EH9 3FD, UK\\
\aff{CERN} Theoretical Physics Department, CERN, 1211 Geneva 23, Switzerland\\
\aff{CU} Physics Department, Columbia University, New York, NY 10027, USA  \\
\aff{KU} Department of Physics and Astronomy, University of Kentucky, Lexington, KY 40506, USA \\
\aff{BU} Peking University, Beijing 100871, China \\
\aff{RBRC} RIKEN/BNL Research Center, Brookhaven National Laboratory, Upton, NY 11973, USA\\
\aff{UR} Universit\"at Regensburg, Fakult\"at f\"ur Physik, 93040 Regensburg, Germany\\
\aff{Soton} Physics and Astronomy, University of Southampton, Southampton SO17 1BJ, UK\\
\aff{Milan} Dipartimento di Fisica, Universita\'a degli Studi di Milano-Bicocca, 20126 Milan, Italy \\
\aff{INFN} INFN, Sezione di Milano-Bicocca, 20126 Milan, Italy \\
}


\bibliography{refs}
\bibliographystyle{JHEP-2}

\end{document}